\documentclass[%
 reprint,pre,
 amsmath,amssymb,
 aps,
]{revtex4-2}
\usepackage{natbib}
\usepackage{url}
\usepackage{graphicx}
\usepackage{dcolumn}
\usepackage{bm}
\usepackage{physics}
\usepackage[T1]{fontenc}
\usepackage{amsmath}
\usepackage{amsfonts}
\usepackage{amsthm}
\usepackage{amssymb}
\usepackage{lipsum}

\usepackage[usenames,dvipsnames]{color}
\usepackage{multirow}
\usepackage{nameref}
\usepackage{soul}
\usepackage[colorlinks=true, urlcolor=blue, linkcolor=blue, citecolor=blue, pdftex]{hyperref}
\usepackage{bm}
\usepackage{braket}
\usepackage{verbatim} 
\usepackage{booktabs}
\usepackage[version=4]{mhchem}
\begin{document}
\title{Ergotropic advantage in a measurement-fueled quantum  heat engine}
\author{Sidhant Jakhar}
\email{ph20036@iisermohali.ac.in}
\author{Ramandeep S. Johal}
\email{rsjohal@iisermohali.ac.in}
\affiliation{Department of Physical Sciences, Indian Institute of Science Education and Research Mohali, Sector 81, SAS Nagar, Manauli PO 140306 Punjab, India}

\begin{abstract}
This paper investigates a coupled 
two-qubits heat engine fueled by generalized measurements of the 
spin components and using a single heat reservoir as sink. 
 {Our model extends the  proposal of 
Yi and coworkers [Phys. Rev. E {\bf 96}, 022108 (2017)]
where the role of a hot reservoir in a four-stroke cycle was  
replaced by a quantum measurement apparatus, the other steps
being two quantum adiabatic strokes and thermalization
with a cold reservoir. We propose a five-stroke cycle, 
where an ergotropy extracting stroke is introduced following
the measurement stroke, and study the effect 
of measurements of different spin components 
on the performance of the machine.} 
For measurements along z-z directions, 
we find two possible occupation distributions that yield an active state and
the ergotropic stroke improves the performance of the engine over the four-stroke cycle. Further, the three-stroke engine ( {without 
the adiabatic strokes}) yields the same performance as
the five-stroke engine. For arbitrary working medium and 
non-selective measurements, we prove that the total 
work output of a five-stroke engine is equal to 
the sum of the work outputs of the corresponding 
four-stroke and  three-stroke engines.
 For measurement directions other than z-z,  
there may be many possible orderings of the post-measurement
probabilities that yield an active state. However,
as we illustrate, for specific cases (e.g. x-x),
a definite ordering may be obtained with the projective measurements.
Thus, we find that the five-stroke engine exploiting
ergotropy outperforms both its four-stroke as well as three-stroke 
counterparts.
\end{abstract}
\maketitle
\newpage
\section{\label{s1}Introduction}
Quantum thermodynamics \cite{Vinjanampathy01102016,Correa2019,mahler2014,benenti2017,
Alicki1979,Kieu2004,allahverdyan2008} 
involves, amongst other pursuits, the study of thermal machines based on a quantum working medium.
The idea of a quantum heat engine was first proposed by Scovil and Schulz-DuBois \cite{scovil1959}, who 
argued that a three-level maser, in simultaneous contact with a hot and a cold reservoir, can be regarded as a heat engine.
State-of-the-art technology now allows us to experimentally control systems at the quantum scale, such as spins \cite{Hubner2014,peterson2019,deassis2019,Ono2020,nettersheim2022,thomas2011,sachin2023}, superconducting qubits \cite{karimi2016,Aamir2025} and trapped ions \cite{Maslennikov2019,Abah2012}. So, the
non-classical resources such as entanglement, quantum coherence, quantum correlations and noise are being hotly pursued for their use in improving the performance of thermal machines away from the classical regime \cite{Hewgill2018,Funo2013,Alicki2013,Hovhannisyan2013,Perarau2015,Korzekwa_2016,Goold_2016,Klatzow2019,Zhang2008,dassonneville2025}. 

In recent years, quantum measurements \cite{Jacobs2014,nielsen2010quantum} have attracted increasing attention  due to their ability to mediate energy exchange—either as work extraction \cite{Opatrny2021} or as effective heat transfer. Measurement-induced energy injection can mimic the role of a hot reservoir, provided the measured observable does not commute with the system Hamiltonian \cite{YiJ2017,DAS2019,Anka2021,Linpeng2024}. Importantly, such engines operate without feedback control, distinguishing them from Szilard-type engines \cite{Szilard1929,Kim2011,Leff2004}. Related developments include quantum refrigerators and cooling protocols \cite{Buffoni2018,elouard2025_1}, weak and continuous measurement schemes \cite{Jacobs2006_1,Allahverdyan2011,ferraz2025}, non-ideal measurements \cite{Panda2023} and work-fluctuation analyses \cite{Debarba2019}. Measurements further induce entropy production, probe–system correlations, and raise questions about associated energetic costs \cite{Jacobs2012, ChandPRE2017, Latune2025}. The classification of measurement-induced energy exchange and its role as a fueling mechanism  \cite{Bresque2021,Serra2022} remains an active research direction.

 {
The main idea of our investigation is that after the 
non-selective quantum measurement injects heat into the 
working medium, the latter may be left in a non-passive or active 
state. From this state, it is possible to extract work by a unitary  process in which the Hamiltonian of the system undergoes a cyclic evolution and the maximum work so extracted is known as the ergotropy \cite{Allahverdyan004}.} 
The present work extends the measurement-fueled engine  \cite{YiJ2017,DAS2019, Behzadi2020} by using weak quantum measurements
on a coupled-qubits system while exploiting the active 
nature of the post-measurement state of the system. 
We  incorporate an ergotropy-extracting stroke \cite{Francica2017,Biswas2022,Touil2022,Hadipour2024,choquehuanca2025} 
after the measurement stroke, thereby upgrading the four-stroke cycle to a five-stroke cycle. 
With measurements on the spin components along 
specific directions, we highlight the thermodynamic 
advantage of the ergotropic stroke by comparing  
the performance of the five-stroke engine
with the four-stroke (without ergotropy)
as well as a three-stroke (without adiabatic strokes) cycle. 
\par \indent
The paper is organized as follows. Section \ref{model} 
describes the five-stroke heat cycle of the measurement-fueled
engine, which is applied to a coupled qubits working medium with z-z measurements in Section \ref{zzcase}, while comparing with the four-stroke and the 
three-stroke cycles in Section \ref{4-stroke} and \ref{3-stroke} respectively. 
Section \ref{xxcase} describes the results for other measurement directions. We conclude our analysis in Section \ref{conclusions}.
\section{Five-stroke engine}\label{model}
\begin{figure}[htp]
    \centering
   \includegraphics[width=0.8\linewidth]{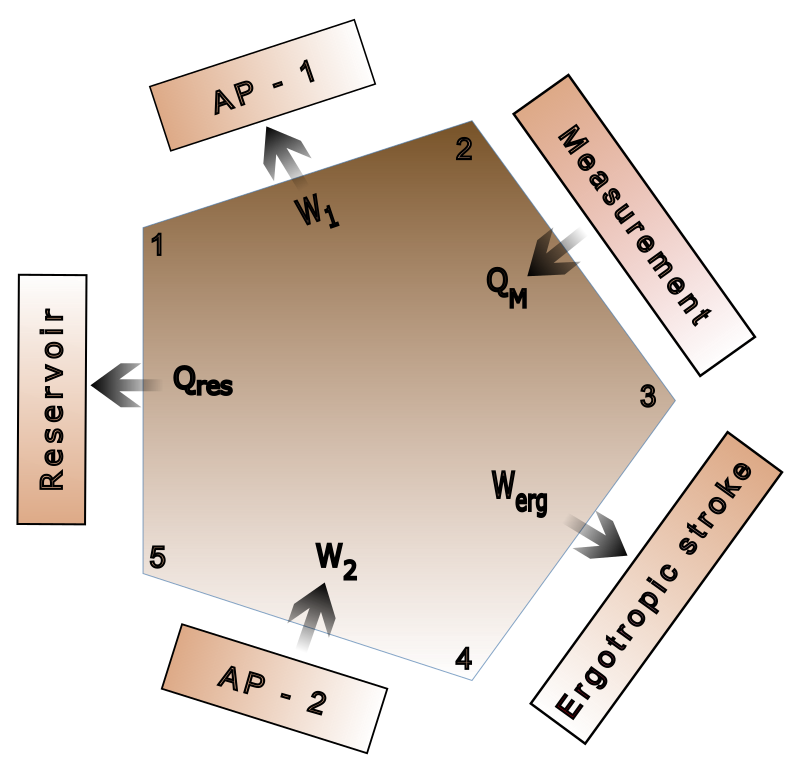}
    \caption{A five-stroke engine cycle consisting of     the first quantum adiabatic stroke ($1\rightarrow2$), the measurement stroke ($2\rightarrow3$), the ergotropy extraction stroke ($3\rightarrow4$), the second quantum adiabatic stroke ($4\rightarrow5$)  and 
    finally the thermalization stroke with the heat 
    reservoir ($5\rightarrow1$). 
     {Without the stroke $3\rightarrow 4$, the cycle
    reduces to the four-stroke engine of Ref. \cite{YiJ2017}}.}
    \label{fig1}
\end{figure}

The working medium, or the system, is described
by the Hamiltonian $H(B)$, with its initial state as the    
thermal state ($\rho^{\rm in}$) corresponding
to a heat reservoir 
at inverse temperature $\beta$, and the control
parameter set at $B=B_2$. 
 {We follow the convention
of the two-reservoir model in which the (magnetic field) parameter
at the cold reservoir is labelled as $B_2$
and in the present case, the only available reservoir 
acts as a cold reservoir.}
The occupation 
probability for the $n^{\rm th}$ energy eigenstate
is labelled as $p_n$. The system undergoes
a heat cycle consisting of five strokes,  as described below (see Fig. \ref{fig1}).
\par\noindent
\text{{\it Stroke} $1 \to 2$}: This is a quantum adiabatic process ({AP-1}) in which the parameter $B$ is increased from $B_2$ to $B_1$ sufficiently slowly such that the quantum adiabatic theorem \cite{Born1928} holds and the occupation probabilities remain unchanged. 
 {Since the system remains in its instantaneous energy eigenstate, the state at
the end of this stroke is given by: 
$\rho = \sum_n p_n |n,B_1 \rangle \langle n, B_1|$, where
$|n,B_1 \rangle$ are energy eigenstates at the end of this process. }
\par\noindent
\text{{\it Stroke} $2 \to 3$}: Keeping the Hamiltonian fixed at $H(B_{1})$, a generalized discrete measurement of an observable 
${\cal O}$ (incompatible with $H(B_{1})$) is made. This measurement changes the state of the system, with the  post-measurement occupation probabilities denoted as $p^{\rm{\rm PM}}_{n}$.
\par\noindent
\text{{\it Stroke} $3\to 4$}: 
 {This is termed as the ergotropic stroke.
Following Ref. \cite{Allahverdyan004},
the ergotropy is defined as the maximum work extractable from a quantum system
via a cyclic unitary on its Hamiltonian. The state of the 
system from which work can be so extracted is denoted as an active
state, otherwise it is called a passive state.
Thus, the stroke $3\to 4$ is executed only if the post-measurement state is 
an active state  (the criterion is explained in the next Section). 
In other words, if the post-measurement state is a passive state,
then this stroke is not executed and the proposed heat cycle 
reduces to the four-stroke cycle of Ref. \cite{YiJ2017},
with all the other steps remaining the same.}   \\
\text{{\it Stroke} $4\to 5$}: This is again a quantum adiabatic process (AP-2) in which the system is isolated and the parameter $B$ is slowly brought back to its initial value $B_{2}$. 
\par\noindent
\text{{\it Stroke} $5 \to 1$}: Finally, the system thermalizes with the heat reservoir and returns to the initial state, thus 
completing the cycle. 

The exchange of heat or work in a stroke is defined 
as the difference of the final and the initial mean energy of the system during that stroke.  
The work in the AP-1 stroke  is given by
\begin{equation}
    W_{1} = 
    \sum_{n} [E_{n}(B_{1})-E_{n}(B_{2})]p_{n}.
    \label{w1}
\end{equation}
The generalized measurement is characterized by the set of  hermitian operators ($M_{\alpha}=M_{\alpha}^{\dagger}$) which satisfy the completeness relation $\sum_{\alpha}M_{\alpha}^{}M_{\alpha}^{\dagger} = I$.
The condition that these operators do not commute with the Hamiltonian $H(B_{1})$ ensures that the measurement process  
adds energy (see Eq. (\ref{qm})) to the system in the form of heat \cite{YiJ2017}. 
For a non-selective measurement where the outcomes are not 
recorded, the post-measurement state
takes the form: $\rho^{\rm{PM}} = \sum_{\alpha}M_{\alpha}\rho M^{\dagger}_{\alpha}$, with occupation probabilities over energy eigenstates are given by  
$p^{\rm{PM}}_{n} =\bra{n,B_1}\rho^{\rm{PM}}\ket{n,B_1}$. 
The heat absorbed by the system in the measurement stroke is then 
\begin{equation}
     Q_{\rm M} =  
     \sum_{n}E_{n}(B_{1})[p^{\rm{\rm PM}}_{n}-p_{n}] >0.
     \label{qm}
\end{equation}
  {To implement the ergotropic stroke, suppose that under the action of 
a cyclic unitary $U$, the post-measurement state $\rho^{\rm{PM}}$
transforms to $U\rho^{\rm PM} U^{\dag}$ while
the Hamiltonian returns to its original form $H(B_1)$. The average 
work extracted in this process is given by 
$W(\rho^{\rm{PM}}, U) = 
{\rm Tr}[H(U \rho^{\rm{PM}} U^{\dag} - \rho^{\rm{PM}})] < 0$. 
The maximum of $W(\rho^{\rm{PM}}, U)$ over all cyclic unitaries
is called the ergotropy $W_{\rm erg}$. 
After the extraction of ergotropy (stroke $3 \to 4$),
the passive state is given by $\rho' = \sum_n p_n' \ket{n,B_1} \bra{n,B_1}$, where due to the 
unitary nature of the ergotropic stroke, 
$p_n'$ are just the shuffled eigenvalues of $\rho^{\rm PM}$ such that $p_n' > p_{n+1}'$ for $E_n(B_1) < E_{n+1}(B_1)$.
} 
Then, the ergotropy is given by
\begin{equation}
    W_{\rm{\rm erg}} = 
    \sum_{n}E_{n}(B_{1})[p'_{n}-p^{\rm{PM}}_{n}] < 0.
    \label{werg}
\end{equation}
Next, the work in the AP-2 stroke  is
\begin{equation}
    W_{2} = 
    \sum_{n}[E_{n}(B_{2})-E_{n}(B_{1})]p_{n}'.
    \label{w2}
\end{equation}
Finally, the heat rejected to the reservoir is 
\begin{equation}
    Q_{\rm res} = 
    \sum_{n}E_{n}(B_{2})[p_{n}-p'_{n}] < 0.
    \label{qc}
\end{equation}
The sign of $Q_{\rm res}$ may be argued on 
thermodynamic grounds as follows. The state of the 
working medium or the system undergoes a cycle.
The measurement step causes an increase in 
the entropy of the system while the adiabatic
strokes keep the entropy unchanged. So, in order to 
return to initial state, the system
must release heat/entropy to the reservoir, 
implying that $Q_{\rm res} < 0$. 

Using energy conservation, the total work extracted in a cycle is
\begin{equation}
    W_{\rm T}^{(5)} = -W_1 -W_2 - W_{\rm erg} = 
    Q_{\rm M}+Q_{\rm res}.
    \label{wt5}
\end{equation}
The operation of the engine requires  
$W_{\rm T}^{(5)} > 0$.  
So, the efficiency is defined as
\begin{equation}
    \eta  = \frac{W_{\rm T}^{(5)}}{Q_{\rm M}}.
\end{equation}
In the next section, we apply the above
framework to a coupled qubits working medium where we consider spin measurements on each qubit, along specific directions. 
\section{\label{coupled qubits}Coupled-qubits working medium}
We consider two coupled qubits following a 1D Hamiltonian with
isotropic Heisenberg interaction \cite{arnesen2001}: 
\begin{equation}
H(B) = B(\sigma_{z}^{\rm A}\otimes I^{\rm B}+I^{\rm A}\otimes \sigma_{z}^{\rm B})
+ 2J\!\sum_{i=x,y,z} \sigma_{i}^{\rm A}\otimes \sigma_{i}^{\rm B}.    
\end{equation}
The magnetic field ($B$), applied along the z-axis, is the control 
parameter whereas the coupling strength $J>0$
(anti-ferromagnetic case) is held fixed during the cycle.
$I^{\rm A/B}$ and 
$\sigma_{i}^{\rm A/B}$ are  respectively 
the identity operator and Pauli matrices of 
qubit A or B. 
We study the engine in the 
strong-coupling regime ($4J>B$) so that
 the energy eigenvalues in ascending order are 
$E_1=-6J, E_2 =2J-2B, E_3=2J$ and $E_4=2J+2B$, with $(\ket{10}-\ket{01})/\sqrt{2} =\ket{\psi_{-}}$, $\ket{11}$, $(\ket{10}+\ket{01})/\sqrt{2}=\ket{\psi_{+}}$ and $\ket{00}$ as the corresponding eigenstates. 
 {We choose to study this model
in the regime $4J > B$, since it
does not perform as an Otto engine with 
two-reservoirs set up in this regime \cite{Altintas2015}. 
However, the measurement
based four-stroke  engine performs with a general choice of directions
and so provides a potential alternative to the standard 
two-reservoir engine \cite{DAS2019}.}

Now, the initial state is  $\rho^{\rm{in}} = {e^{-\beta H(B_{2})}}/{Z}$, where $Z = \sum_{n=1}^{4} e^{-\beta E_{n}}$ is the partition function. Then, we have 
$p_{n} = e^{-\beta E_{n}}/Z$.
In the first adiabatic stroke, the work [Eq. (\ref{w1})] 
is evaluated to be 
\begin{align}
    W_{1} = & -2(B_{1}-B_{2})\left(p_{2}-p_{4}\right) <0. 
\end{align}

We choose $M_{\alpha}$ operators in arbitrary directions
$\hat{n}^{\rm A}$ 
and $\hat{n}^{\rm B}$,   
in the following form: 
\begin{align}\label{Mab}
\begin{split}
      M_{\pm,\pm} & = (c_{0}I^{\rm A} \pm c_{1}\vec{\sigma}^{\rm A}.\hat{n}^{\rm A})\otimes (c_{0}I^{\rm B} \pm c_{1}\vec{\sigma}
      ^{\rm B}.\hat{n}^{\rm B}) \\
   M_{\pm,\mp} & = (c_{0}I^{\rm A} \pm c_{1}\vec{\sigma}^{\rm A}.\hat{n}^{\rm A})\otimes (c_{0}I^{\rm B} \mp c_{1}\vec{\sigma}^{\rm B}.\hat{n}^{\rm B})
\end{split}   
\end{align}
where $c_0$ and $c_1$ are real parameters.
From the completeness relation: 
$2c_{0}^{2} + 2c_{1}^{2} = 1$.  
One of the parameters, say $c_0$, can be chosen to
define the strength of the measurement. Thus, 
$c_0=c_1=1/2$ implies a strong or projective measurement \cite{DAS2019}.
The general expressions for the post-measurement probabilities are quite complicated and it is not feasible to 
determine the conditions for their relative ordering. Now, we study measurements along specific 
directions. 
\subsection{\label{zzcase}z-z measurement}
We choose 
$\vec{\sigma}^{\rm A}.\hat{n}^{\rm A} = \sigma_{z}^{\rm A}$
and $\vec{\sigma}^{\rm B}.\hat{n}^{\rm B} = \sigma_{z}^{\rm B}$. The post-measurement occupation probabilities 
are given by
\begin{align}\label{59}
\begin{split}
    p_{1}^{\rm{PM}} & = p_{1} - 4c_{0}^{2}(1-2c_{0}^{2})(p_{1}-p_{3}), \\
    p_{2}^{\rm{PM}} & = p_{2},\\
    p_{3}^{\rm{PM}} & = p_{3} + 4c_{0}^{2}(1-2c_{0}^{2})(p_{1}-p_{3}),\\
    p_{4}^{\rm{PM}} & = p_{4}.
    \end{split}
\end{align} 
The heat absorbed by the system during the measurement stroke [Eq. (\ref{qm})] is 
\begin{equation}
  Q_{\rm M} = 8 J (p^{\rm{PM}}_{3}-p_{3}) > 0.
   \label{qm2bit}
\end{equation}
As a result of the measurement,
 the occupation probabilities 
of levels $E_2$ and $E_4$ are unchanged, while
those of $E_1$ and $E_3$ come closer to 
each other. Intuitively, we see that the post-measurement
probability distribution becomes more uniform and so 
its Shannon entropy increases.
Also, 
$p^{\rm{PM}}_{1}-p^{\rm{PM}}_{3} = (4c_{0}^{2}-1)^{2}(p_{1}-p_{3}) 
\geq 0$, where the equality is obtained for $c_0=1/2$.
Note that $p_4$ is still the lowest probability in 
the distribution (\ref{59}) since the 
probabilities were initially ordered as $p_{1}>p_{2}>p_{3}>p_{4}$. Based on these considerations, we can infer: 
$p^{\rm{PM}}_{1} \geq p^{\rm{PM}}_{3} > p_{4}$.
 {Note that $\rho^{\rm PM}$ has no coherence 
in the energy eigenbasis for the z-z case.
Therefore, to ascertain the active nature of 
the state $\rho^{\rm{PM}}$,
we  have to determine the relative ordering of 
the probabilities $p^{\rm{PM}}_{n}$}.  
It can be seen that  an active post-measurement 
state implies one of the
following two possibilities:
\begin{align*}
  ({\rm R1} ) \qquad & p^{\rm{PM}}_{1} \geq p^{\rm{PM}}_{3} > p_{2} > p_{4}, \\
     ({\rm R2} ) \qquad & p_{2} > p^{\rm{PM}}_{1} \geq p^{\rm{PM}}_{3} > p_{4}.
\end{align*}
\begin{figure}
\centering
    \includegraphics[scale=0.472]
    {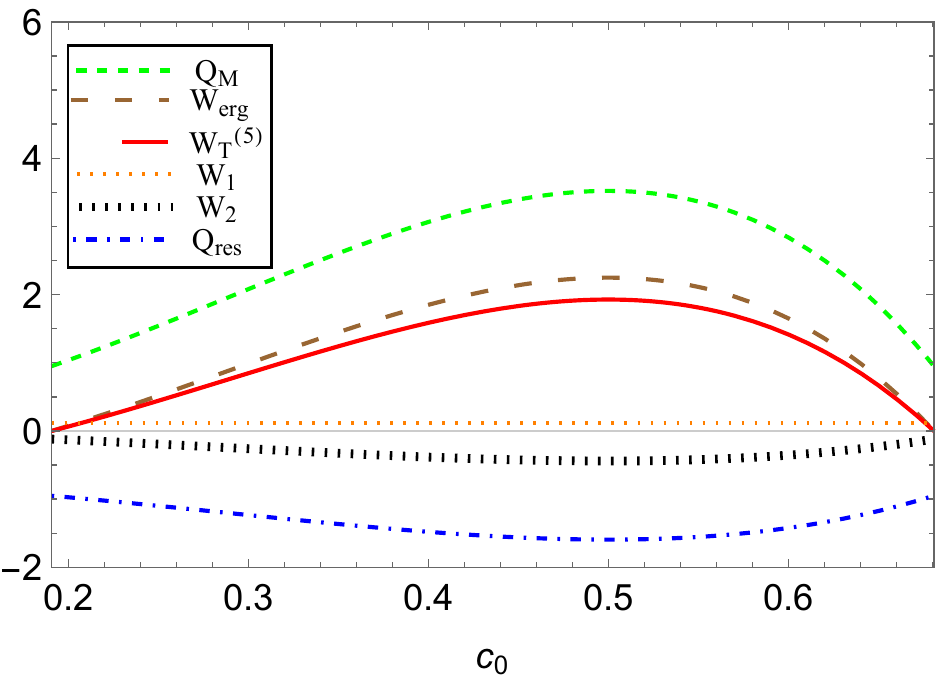}
    \caption{R1 case: Heat and work contributions in the five-stroke cycle vs. $c_{0}$. Parameters are set at $B_{1} = 3.5$, $B_{2}$ = 3, $J$ = 1 and $\beta$ = 1.}
    \label{fig2}
\end{figure}
Let us consider the R1 case
in detail. Now, in the ergotropy stroke, the occupation probabilities $p^{\rm PM}_{n}$ get reordered in an opposite sense 
to the energy eigenvalues \cite{Allahverdyan004}, implying that 
$p_1' = p^{\rm{PM}}_{1}$, $p_2' = p^{\rm{PM}}_{3}$, $p_3' = p_2$
and $p_4' = p_4$. 
Thus, the  system state {\it after} the ergotropy stroke can be given as
\begin{align} 
    \rho'  = &
~p^{\rm{PM}}_{1}\ket{\psi_{-}}\bra{\psi_{-}} + p^{\rm{PM}}_{3}\ket{11}\bra{11} + 
    p_{2}\ket{\psi_{+}}\bra{\psi_{+}} 
    \nonumber \\
    &+ p_{4}\ket{00}\bra{00}.
\end{align}
Note that R1 case is characterized by the condition
$p^{\rm{PM}}_{3} > p_{2}$, which can be 
cast in the form:
\begin{equation}
    4c_{0}^{2}(1-2c^{2}_{0}) > 
    \frac{e^{2\beta B_2}-1}{e^{8\beta J}-1} = \chi (B_2, J, \beta),
\end{equation}
which implies that R1 case bounds the 
parameter $c_0$ as follows:
$(1 - \sqrt{1-2\chi})/4 < c_{0}^{2} 
< (1 + \sqrt{1-2\chi})/4$, where $\chi < 1/2$. 

\begin{figure}[!h]
    \centering
    \includegraphics[scale=0.518]{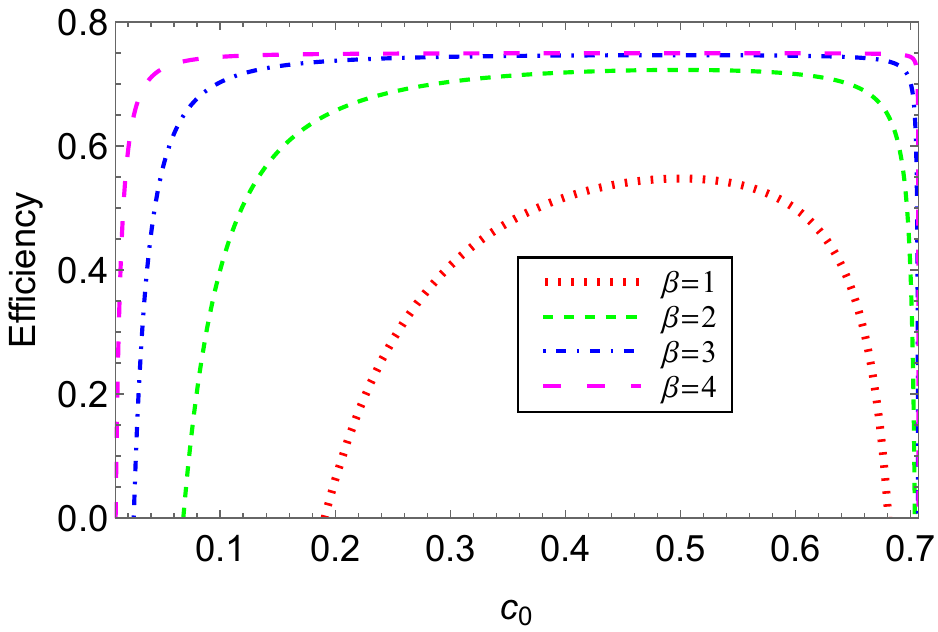}
    \caption{R1 case: Efficiency [Eq. (\ref{eqn89})]  vs. $c_{0} \in [0,1/\sqrt{2}]$. $B_{2} = 3$ and $J = 1$. So, the limiting value of the efficiency (as 
    $\beta \to \infty$) is $B_2/4J = 0.75$.}
    \label{fig3}
\end{figure}

\begin{figure}[!h]
    \centering
    \includegraphics[scale=0.529]
    {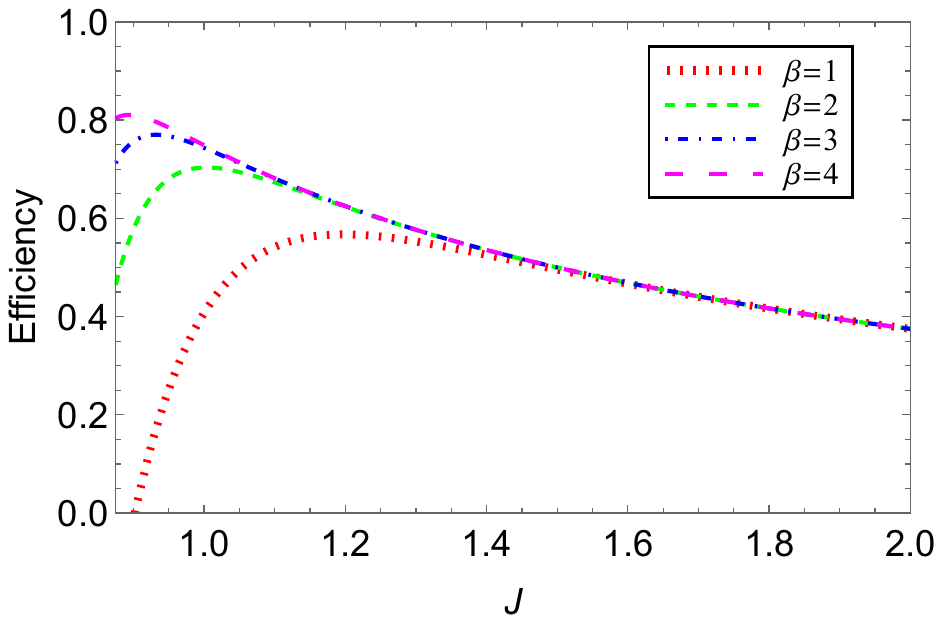}
    \caption{R1 case: Efficiency [Eq. (\ref{eqn89})] vs. $J$. $B_{2} = 3$ and $c_{0}=0.3$.}
    \label{fig4}
\end{figure}

Next, the ergotropy [Eq. (\ref{werg})] is given by
\begin{equation}
    W_{\rm{erg}} = -2B_{1}(p^{\rm{PM}}_{3}-p_{2})  <0.
\end{equation}
Work in the second adiabatic stroke [Eq. (\ref{w2})] is 
\begin{align}
    W_{2} &= 2(B_{1}-B_{2})\left(p^{\rm{PM}}_{3}-p_{4}\right) >0.
\end{align}
Finally, the heat exchange in 
the thermalization stroke [Eq. (\ref{qc})] is
\begin{align}
    Q_{\rm res} = 2B_2(p^{\rm{PM}}_{3}-p_{2}) -8J 
    (p^{\rm{PM}}_{3}-p_{3}). 
\end{align}
Using Eq. (\ref{59}), we can show
that $Q_{\rm res} < 0$, as required. 
The net work extracted is  
\begin{equation}
    W_{\rm T}^{(5)} = Q_{\rm M} + Q_{\rm res} = 2B_{2}\left(p^{\rm{PM}}_{3}-p_{2}\right) > 0.
    \label{WT5}
\end{equation} 
In Fig. \ref{fig2}, the above heat and work contributions
are plotted for a specific example. 
It is observed that the work output 
becomes optimal at $c_{0}$ = 1/2
i.e. when  $p^{\rm{PM}}_{3}$ 
achieves its maximum value $(p_1+p_3)/2$. The efficiency in the R1 case can be expressed as
\begin{align}
     \eta =  & \frac{B_{2}}{4J}\left[1 -\frac{(e^{2\beta B_{2}}-1)}{4c_{0}^{2}(1-2c_{0}^{2})\left(e^{8\beta J}-1\right)}\right].
\label{eqn89}           
\end{align}
It is interesting to note that  $\eta$ is independent of the parameter $B_{1}$.
This is not a generic feature,
but depends on the energy spectrum.
For the given model,
$\eta$ is maximum at $c_0 = 1/2$ i.e. for 
projective measurements. 
The efficiency is plotted in Fig. \ref{fig3}.
For a given $c_0$, 
$\eta$ increases with $\beta$. 
 For low enough temperatures, $\eta$ grows rapidly  with $c_0$ and approaches its limiting value $B_{2}/4J$. 
 In Fig. \ref{fig4}, the efficiency is plotted versus $J$  for a fixed $c_{0}$ at various temperatures. Again, decreasing the temperature increases the efficiency at low $J$ values,
 while for high $J$ values, all efficiency curves 
merge into each other and goes to zero.

Similarly, we can study the R2 case
which is characterized by the ordering: 
$p_{2} > p^{\rm{PM}}_{1} \geq p^{\rm{PM}}_{3} > p_{4}$. 
To appreciate how this ordering may come about, consider the difference:
   $p_{2} - p^{\rm{PM}}_{1} = \left(p_{2}-p_{1}\right) + 4c_{0}^{2}(1-2c^{2}_{0})\left(p_{1}-p_{3}\right)$.
As we are working in the regime $4J > B_2$, so $B_{2}$ close to $4J$ implies that the lowest two levels
are close to each other. Then, the term $(p_{2}-p_{1})$ will be negligible and the condition $p^{}_{2} > p^{\rm PM}_{1}$ can prevail, thereby producing the R2 case. 

\begin{figure}[!h]
    \centering
    \includegraphics[scale=0.475]{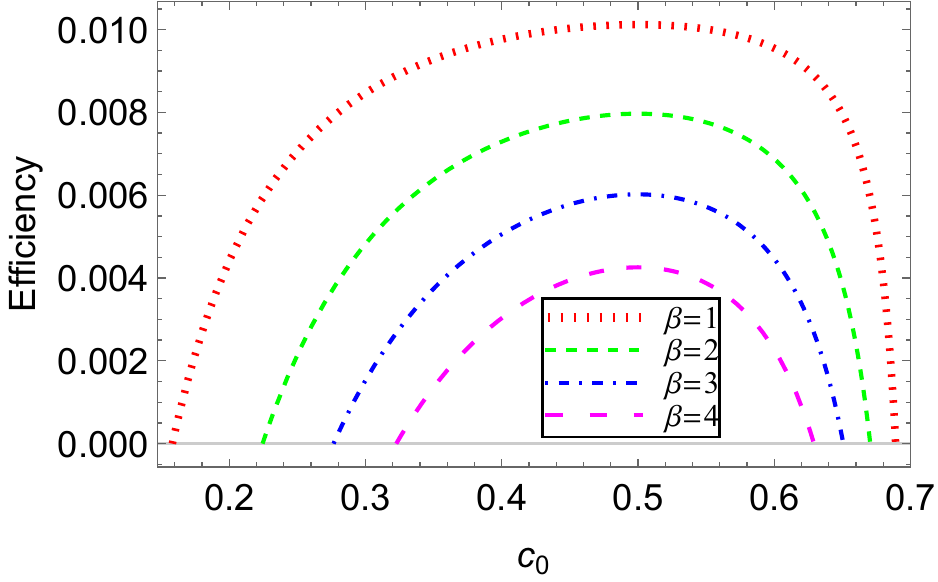}
    \caption{R2 case: Efficiency vs. $c_{0}$. $B_{2} = 3.9$, $J = 1$. 
    The efficiency decreases as the temperature is lowered at
    a given measurement strength, unlike the R1 case (see Fig. \ref{fig3}).
    The R2 efficiency is also maximized at $c_0=1/2$.}
    \label{fig5}
\end{figure}

The total work output is 
\begin{align}
   W_{\rm T}^{(5)} = & 2(4J-B_{2})\left(p_{2}-p^{\rm{PM}}_{1}\right) > 0,
   \end{align}
   whereas $Q_{\rm M}$ is given by Eq. (\ref{qm2bit}). 
   The efficiency of this case can then be computed. 
As shown in  Fig. \ref{fig5}, the R2 case yields a rather small efficiency. Another notable feature is that upon decreasing the temperature, the efficiency decreases, thus showing an opposite trend to the R1 case.

\subsection{\label{4-stroke}Four-stroke cycle}
For comparison, we also consider the four-stroke cycle \cite{YiJ2017, DAS2019} 
that does not involve the 
ergotropic stroke. 
It implies the occupation probabilities in AP-2 are given by 
$p^{\rm{PM}}_{n}$. The expressions for $W_2$ and $Q_{\rm res}$ are obtained by replacing  $p_n'$ with $p^{{\rm PM}}_{n}$ in the corresponding expressions for the five-stroke cycle as in Sec. \ref{model}.
Thus, the total work extracted can be expressed as:
\begin{align}
    W_{\rm T}^{(4)} = \sum_{n}[E_{n}(B_{2})-E_{n}(B_{1})](p_n -p_{n}^{\rm PM}).
    \label{wt4}
\end{align}
For the two-qubits system and with z-z measurements, we obtain that the net work 
output of the four-stroke cycle 
is zero. This feature was noted in Ref. \cite{DAS2019}
for z-z projective measurements and we affirm 
it for generalized measurements too. 
Thus, we can conclude that the ergotropy stroke 
improves the performance of the engine based on 
generalized z-z measurements. 
\subsection{\label{3-stroke}Three-stroke cycle}
As an alternative, we consider an ergotropy based
three-stroke cycle
consisting of the following steps.
We prepare the system with Hamiltonian $H(B_2)$ in a 
thermal state $\rho^{\rm in}$ at inverse temperature $\beta$.
i) The system is isolated from the reservoir 
(assuming a weak interaction with the reservoir and
 so negligible costs in attaching and detaching the system from the reservoir) and nonselective generalized measurements 
 are performed which transform the state 
to $\rho^{\rm PM}$. Heat transferred due to
measurement is given by
\begin{equation}
     Q_{\rm M} =  
     \sum_{n}E_{n}(B_{2})[p^{\rm{\rm PM}}_{n}-p_{n}] >0.
     \label{qm3}
\end{equation}
ii) 
Ergotropy, $W_{\rm erg}$, 
is extracted which yields the passive state
$\rho'$. iii) Finally, the system is brought to the initial 
state by thermalization with the reservoir. The net
work extracted, equal in magnitude to the ergotropy 
in step (ii), is given by 
\begin{align}
W_{\rm T}^{(3)} = -W_{\rm erg} =
\sum_{n} E_{n}(B_{2})(p_{n}^{\rm PM}-p_n').
\label{wt3}
\end{align}
From Eqs. (\ref{wt5}), (\ref{wt4}) and (\ref{wt3}), we
obtain an interesting equality:
\begin{equation}
    W_{\rm T}^{(5)} = W_{\rm T}^{(4)} + W_{\rm T}^{(3)}.
    \label{wt543}
\end{equation}
Note that the above relation holds for 
arbitrary Hamiltonian and nonselective 
measurements.
It implies that the performance of a
four-stroke measurement engine ($W_{\rm T}^{(4)} \geq 0$),
with an active post-measurement state, 
can be enhanced by using the five-stroke cycle.

For the R1 case, we obtain 
$W_{\rm T}^{(3)} = 2B_{2}(p^{\rm{PM}}_{3}-p_{2})$,
which, as expected, is equal  to the work
in the five-stroke cycle [Eq. (\ref{WT5})] 
since $W_{\rm T}^{(4)} =0$ for z-z measurements. 
$Q_{\rm M} = 8 J (p^{\rm{PM}}_{3}-p_{3}) > 0$ for the three-stroke cycle, which is the same as Eq. (\ref{qm2bit}). Therefore, its efficiency is  
the same as for the five-stroke cycle [Eq. (\ref{eqn89})]
 {in the case of z-z measurements.} 
A similar conclusion holds for the case R2. 
\section{\label{xxcase} Other measurement directions}
We also studied the performance of the engine with 
generalized measurements along x-x (y-y), x-y, and x-z (y-z) directions. 
 {In general, 
the post-measurement state $\rho^{\rm PM}$
contains some coherence (in the energy basis), making
it an active state. Then, to determine
the passive state after the ergotropic stroke, 
we need to order the 
eigenvalues of the state $\rho^{\rm PM}$ 
in a non-increasing sense.} To illustrate,
we consider x-x measurements for which we set 
$\vec{\sigma}^{\rm A}.\hat{n}^{\rm A} = \sigma_{x}^{\rm A}$
and $\vec{\sigma}^{\rm B}.\hat{n}^{\rm B} = \sigma_{x}^{\rm B}$ in Eq. (\ref{Mab}).
For the case of x-x weak measurements,
the state $\rho^{\rm PM}$ is active, and 
we get the post-measurement occupation 
probabilities as
\begin{align}\label{xxpm}
\begin{split}
p^{\rm{PM}}_{1} &= 4c_{0}^{4}p_{1}+4c_{1}^{4}p_{1} + 4c_{0}^{2}c_{1}^{2}(p_{2}+p_{4}),\\
p^{\rm{PM}}_{2} &= 4c_{0}^{4}p_{2}+4c_{1}^{4}p_{4} + 4c_{0}^{2}c_{1}^{2}(p_{1}+p_{3}),\\
p^{\rm{PM}}_{3} &= 4c_{0}^{4}p_{3}+4c_{1}^{4}p_{3} + 4c_{0}^{2}c_{1}^{2}(p_{2}+p_{4}),\\
p^{\rm{PM}}_{4}&= 4c_{0}^{4}p_{4}+4c_{1}^{4}p_{2} + 4c_{0}^{2}c_{1}^{2}(p_{1}+p_{3}).\\
\end{split}
\end{align}
The relative ordering of the above probabilities 
is not straightforward, though we can prove specific inequalities, such  as $p^{\rm{PM}}_{1}>p^{\rm{PM}}_{3}$. 
 For $c_{0} < 1/2$, $p^{\rm{PM}}_{2} < p^{\rm{PM}}_{4}$  while $p^{\rm{PM}}_{2} > p^{\rm{PM}}_{4}$ for $c_{0} > 1/2$. 
Unlike the case of z-z measurements, here 
the four-stroke cycle  with weak measurements
yields a non-zero total work.

{The eigenvalues of $\rho^{\rm PM}$ 
can be written as 
\begin{align}
\begin{split}
    p_1' & =    p^{\rm{PM}}_{1}, \\
    p_2' & =  \frac{p^{\rm{PM}}_{2} + p^{\rm{PM}}_{4}}{2} +\Delta,  \\
    p_3' & = p^{\rm{PM}}_{3}, \\
    p_4' & = \frac{p^{\rm{PM}}_{2} + p^{\rm{PM}}_{4}}{2} -\Delta, 
\end{split}
\label{pnpxxw}
\end{align}
where $\Delta>0$ depends in a complicated way on $p_n$ and $c_0$. 
The above eigenvalues are also 
the occupation probabilities in the passive state
$\rho'$, to be ordered in a decreasing sense.
However, there can be different orderings
of these probabilities in different parameter
regimes. 
On the other hand, for the specific case of projective measurements ($c_0=1/2$),
the post-measurement probabilities are simplified as:
$p^{\rm{PM}}_{1} = (1+p_1-p_3)/4, \; 
p^{\rm{PM}}_{2} = p^{\rm{PM}}_{4} = 1/4, \;
p^{\rm{PM}}_{3} = (1-p_1+p_3)/4$, 
while the eigenvalues of $\rho^{\rm PM}$ are 
$p_1' = p_2' =   p^{\rm{PM}}_{1}$ and 
$p_3' = p_4' = p^{\rm{PM}}_{3}$,
which are already ordered  
i.e. $p_1' = p_2' > p_4' = p_3'$, since 
$p^{\rm{PM}}_{1} > p^{\rm{PM}}_{3}$.
In this special case, the heat exchanged during the measurement stroke is calculated to be 
\begin{align}
    Q_{\rm M} = 2 B_{1}(p_{2}-p_{4}) + 2J(2p_{1}-p_{2}-p_{4}).
    \label{qmxx_proj}
\end{align} 
The ergotropy is given by $W_{\rm erg} = -B_{1}(p_{1}-p_{3}) < 0$, while the total work output is
\begin{align}
    W_{\rm T}^{(5)} = 2(B_{1}-B_{2})(p_{2}-p_{4}) + B_{2}(p_{1}-p_{3})
    > 0.
    \label{W total for xx five-stroke projective}
\end{align}
The work output of a four-stroke cycle 
with projective measurements is 
$W_{\rm T}^{(4)}=2(B_{1}-B_{2})(p_{2}-p_{4}) < W_{\rm T}^{(5)}$, 
while the three-stroke cycle yields
$W_{\rm T}^{(3)} = {B_{2}}(p_{1}-p_{3})$.
Thus, we verify the equality (\ref{wt543}).}  

\begin{figure}
    \centering
    \includegraphics[scale=0.53]{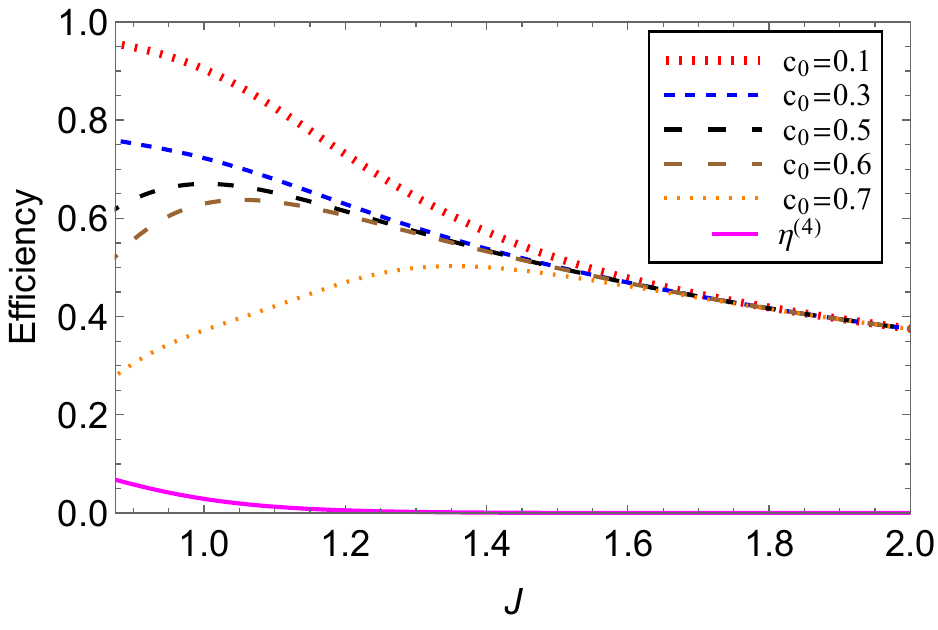}
    \caption{x-x case: Efficiency vs. $J$ for the five-stroke engine. Weak measurements with $c_0 < 1/2$ can lead to a higher efficiency than with projective measurements ($c_0=1/2$). For comparison, the efficiency of the four-stroke engine with projective measurements, $\eta^{(4)}$, is also 
    plotted. Here, $B_{1}=3.5$, $B_{2}=3$ and $\beta = 1$.}
    \label{fig6}
\end{figure}

In Fig. \ref{fig6}, we show the effect of 
weak measurements 
as they can lead to a higher efficiency compared to 
projective measurements. Here, the ordering 
of the probabilities in the post-ergotropy state
is performed numerically for the given choice
of system parameters.
%
Similarly, we find that the post-measurement state
for the x-z case has coherence and so is an active state.
On the other hand, with x-y projective 
measurements, we find that the post-measurement probability 
distribution is uniform ($p_{n}^{\rm PM} =1/4$). 
Being a passive state, the ergotropic stroke does not help in this case.

Before closing this section, we consider the case 
of spin-measurements along arbitrary directions. 
As mentioned earlier, ordering relations between the post-measurement
probabilities or between the eigenvalues of $\rho^{\rm PM}$ 
are hard to obtain here. However,
if we restrict ourselves to projective measurements, 
the following inequality always holds.
\begin{align}
    p^{\rm{PM}}_{2}-p^{\rm{PM}}_{4} = & \frac{1}{2}(\cos^{2}{\theta_{\rm A}} + \cos^{2}{\theta_{\rm B}})(p_{2}-p_{4}) \geq 0. 
\end{align}
The equality $p^{\rm{PM}}_{2} = p^{\rm{PM}}_{4}$ 
is recovered for projective measurements in $xy$-plane for each qubit ($\theta_{\rm A} = \theta_{\rm B} = \pi/2$) 
irrespective of the 
angles $\phi_{\rm A}$ and $\phi_{\rm B}$. On the other hand, for z-z measurements 
 ($\theta_{\rm A}$ = $\theta_{\rm B}$ = 0), 
 we obtain $p^{\rm{PM}}_{2} = p_2$ and 
 $p^{\rm{PM}}_{4} = p_4$, as found in Section \ref{zzcase}.  
\section{\label{conclusions}Conclusions}
In this paper,  {we studied 
the effect of additional ergotropy extraction 
on the performance of 
a measurement-fueled quantum engine based on a coupled two-qubits system.} Here, the weak measurements of the spin components are used as a resource for input heat. We first analyzed the case of z-z weak measurements in a five-stroke cycle. 
The corresponding four-stroke cycle yields zero work output.
Notably, the post-measurement state has no coherence
in the energy eigenbasis. 
However, there are two possible orderings of the post-measurement probabilities leading to  an active post-measurement state. We find that an ergotropy-extracting stroke inserted after the measurement stroke enables a net positive work output. Furthermore, a reduced three-stroke cycle (without the adiabatic strokes) yields the same performance as the five-stroke cycle. However, for other measurement directions (e.g., x–x, x–z), the post-measurement active state,
in general, has coherence. 
Here, the four-stroke engine already delivers non-zero work, yet the inclusion of an ergotropic stroke enhances the performance. We also find that weak x–x measurements can surpass the projective ones in efficiency, highlighting the role of the measurement strength as a tuning parameter. In contrast, some settings (e.g., x–y case with projective measurements) yield passive post-measurement states, offering no ergotropic advantage. 

A general identity, valid for arbitrary Hamiltonian and non-selective measurements, was established,   which shows that the total work of the five-stroke engine equals the sum of its four-stroke and three-stroke counterparts.
The present findings underline three key points: (i) measurements combined with ergotropy extraction can increase useful work; (ii) engine performance depends sensitively on the measurement strength; and (iii) appropriate choices of these parameters can outperform both standard four-stroke and three-stroke cycles.

 {In our analysis, we have neglected some of the energetic and thermodynamic 
costs towards the implementation of the unitary protocols 
\cite{Abah2019, Deffner2021}.  
The notion of work extraction
from a quantum system assumes some coupling with a work 
reservoir, or a quantum battery. 
In certain situations, with a battery of a finite size and a finite coherence, the work extracted is expected be less than ergotropy \cite{Monsel2020}. 
One possibility is to assume that the battery
is large enough and highly coherent, so that 
all the energy transferred 
to the battery can be interpreted as work. Further,
the post-measurement state of the working medium 
 may have some degree of 
 coherence. Then, extracting the coherent part
 of the ergotropy also becomes an experimental challenge.
 We have assumed a quasi-static heat cycle. Thus, in the present framework, we take the costs incurred
 through these time-dependent protocols \cite{Lili2025} to 
 be negligible. However, in the corresponding
 finite-time thermodynamic cycle, these costs
 can be important, and accordingly, the comparison 
 between the performance of four-stroke and five-stroke
 engines has to be reassessed.
 } 


In conclusion, we have demonstrated that combining measurement-induced population reshaping with the extraction of ergotropy provides a viable thermodynamic advantage for quantum engines.
 {These features can be exploited in feasible experimental 
platforms such as  NMR \cite{Serra2022} and superconducting circuits \cite{Lili2025} 
which have so far formulated state of the art
protocols for single qubit cases.} 
 Future work can include many-body systems, feedback-based protocols and bounding the performance due to finite-time effects. 
\begin{acknowledgments}
S.J. thanks Sachin Sonkar for useful discussions and acknowledges financial support in the form of Senior Research Fellowship from the Indian Institute of Science education and Research Mohali.
\end{acknowledgments}
%

\end{document}